\documentclass[letterpaper, 10 pt, conference]{ieeeconf}  

\IEEEoverridecommandlockouts      

\usepackage{amssymb}
\usepackage{graphicx}
\usepackage{verbatim}
\usepackage{amsmath}
\usepackage{theorem}
\usepackage{amsfonts}
\usepackage{mathrsfs}
\usepackage{amstext}
\usepackage{color}
\usepackage{url}

\newcommand{\be}[1]{\begin{equation}\label{#1}}
\newcommand{\benon}{\begin{equation*}}  
\newcommand{\bemuln}[1]{\begin{multline}\label{#1}}
\newcommand{\bemul}{\begin{multline*}}
\newcommand{\bee}{\begin{eqnarray*}}
\newcommand{\eee}{\end{eqnarray*}}
\newcommand{\been}[1]{\begin{eqnarray}\label{#1}}
\newcommand{\eeen}{\end{eqnarray}}
\newcommand{\began}[1]{\begin{gather}\label{#1}}
\newcommand{\bega}{\begin{gather*}}
\newcommand{\bealn}[1]{\begin{align}\label{#1}}
\newcommand{\beal}{\begin{align*}}
\newcommand{\bealatn}[2]{\begin{alignat}{#1}\label{#2}}
\newcommand{\bealat}{\begin{alignat*}}
\newcommand{\bexalatn}[1]{\begin{xalignat}\label{#1}}
\newcommand{\bexalat}{\begin{xalignat*}}

\newcommand{\ra}{\rightarrow}

\newcommand{\mb}{\mathbf}

{\theoremstyle{plain}

}
{\theoremstyle{break} \theorembodyfont{\it}
\newtheorem{defi}{Definition}
 }

\def\bd{{\mathbf d}}

\def\bh{{\mathbf h}}

\def\bp{{\mathbf p}}
\def\bq{{\mathbf q}}
\def\br{{\mathbf r}}
\def\bs{{\mathbf s}}

\def\bx{{\mathbf x}}

\def\bI{{\mathbf I}}

\def\bM{{\mathbf M}}

\def\bP{{\mathbf P}}

\def\bS{{\mathbf S}}

\def\texitem#1{\par\smallskip\noindent\hangindent 25pt
               \hbox to 25pt {\hss #1 ~}\ignorespaces}

\newcommand{\scrC}{\mathcal{C}}

\newcommand{\scrE}{\mathcal{E}}

\newcommand{\scrG}{\mathcal{G}}

\newcommand{\scrN}{\mathcal{N}}

\newcommand{\scrS}{\mathcal{S}}

\newcommand{\scrV}{\mathcal{V}}
\newcommand{\scrW}{\mathcal{W}}

\newcommand{\bmu}{\boldsymbol{\mu}}

\global\long\def\calA{\mathcal{A}}

\global\long\def\calH{\mathcal{H}}

\global\long\def\calN{\mathcal{N}}

\global\long\def\calS{\mathcal{S}}

\global\long\def\Tr{\mathrm{Tr}}

\global\long\def\sgn{\mathrm{sgn}}

\begin{document}

\global\long\def\mb#1{\mathbf{#1}}
\global\long\def\ie{\mathit{\mbox{\textit{i.e.}}}}
\global\long\def\bvph{\boldsymbol{\varphi}}

\bibliographystyle{IEEEtran}

\title{Botnet Detection using Social Graph Analysis\authorrefmark{1}
\thanks{Research partially supported by the NSF under grants
CNS-1239021 and IIS-1237022, by the ARO under grants
W911NF-11-1-0227 and W911NF-12-1-0390, and by the ONR under grant
N00014-10-1-0952.}}

\author{Jing Wang\authorrefmark{2} \thanks{$\dagger$ Division of Systems
    Engineering, Boston University,  8 St. Mary's St., Boston, MA 02215,
    {\tt wangjing@bu.edu.}} and Ioannis Ch. Paschalidis\authorrefmark{3}
  \thanks{
  $\ddagger$ Dept. of Electrical \& Computer Eng.,
  Boston University, 8 Mary's St., Boston, MA 02215, 
  {\tt yannisp@bu.edu}.}}
\maketitle


\begin{abstract}
    Signature-based botnet detection methods identify botnets by recognizing
    Command and Control (C\&C) traffic and can be ineffective for botnets that use
    new and sophisticate mechanisms for such communications. To address
    these limitations, we propose a novel botnet detection method that
    analyzes the social relationships among nodes. The method consists of
    two stages: (i) anomaly detection in an ``interaction'' graph among nodes
    using large deviations results on the degree distribution, and (ii)
    community detection in a social ``correlation'' graph whose edges connect
    nodes with highly correlated communications. The latter stage uses a
    refined modularity measure and formulates the problem as a non-convex
    optimization problem for which appropriate relaxation strategies are
    developed. We apply our method to real-world botnet traffic and compare
    its performance with other community detection methods. The results show
    that our approach works effectively and the refined modularity measure
    improves the detection accuracy.
\end{abstract}
\section{Introduction}

A botnet is a network of compromised nodes (bots) controlled by a
``botmaster.'' The most common type is a botnet of network
  computers, which is usually used for Distributed Denial-of-Service
(DDoS) attacks, click fraud and spamming, etc. DDoS attacks comprise
packet streams from disparate bots, aiming to consume some critical
resource at the target and to deny the service of the target to
legitimate clients. In a recent survey, 300 out of 1000 surveyed
businesses have suffered from DDoS attacks and 65\% of the attacks cause
up to \$10,000 loss per hour~\cite{neustar-whitepaper}. Both click fraud
and spamming are harmful to web economy. Click fraud exhausts the
advertisement budgets of businesses in pay-per-click
services~\cite{Daswani2007}, and spamming is popular for malicious
advertisements as well as manipulation of search
results~\cite{Gyongyi2005}.

Because of the huge loss caused by botnets,
detecting them in time is very important. Most of the existing botnet
detection approaches focus on \emph{Command and Control }(C\&C) channels
required by botmasters to command their bots~\cite{Strayer2005,Su2012}.
One mechanism is to filter specific types of C\&C traffic (e.g., ${\tt
  IRC}$
traffic)~\cite{Binkley2006,goebel2007rishi,GuGuofei2008}. Recently,
botnets have evolved to bypass these detection methods by using more
sophisticated C\&C channels, such as $\mathtt{HTTP}$ and $\mathtt{P2P}$
protocols~\cite{Bu2010,Daswani2007}. ${\tt P2P}$ botnets like
Nugache~\cite{lemos2006bot} and Storm worm~\cite{Bu2010} are much harder
to detect and mitigate because they are decentralized. In addition, more
types of C\&C channels are emerging; recent research shows that botnets
start to use ${\tt Twitter}$ as the C\&C channel~\cite{singh2012social}.
It is very challenging to identify and monitor these sophisticated C\&C
channels. Furthermore, the switching cost of C\&C channels is
much lower than the monitoring cost, thus botnet can
bypass detection by changing C\&C channels frequently.



In addition to C\&C channels, botnets have some behavioral characteristics.
First, bots activities are more correlated with each other than normal
nodes~\cite{Abdulla2010,GuGuofei2008}.  Second, bots have more interactions
with a set of \emph{pivotal nodes}, including targets and botmasters.
Compared with C\&C traffic, these behavioral characteristics are harder to
hide.

In this paper, we propose a novel botnet detection framework based on these
behavioral characteristics. Instead of focusing on C\&C channels, we detect
botnets by analyzing the social relationships, modeled as graphs of
nodes. Two types of social graphs are considered: $(i)$ \emph{Social
  Interaction Graphs (SIGs)} in which two nodes are connected if there
is interaction between them, and $(ii)$ \emph{Social Correlation Graphs
  (SCGs)} in which two nodes are connected if their behaviors are
correlated. We apply our method to real-world botnet traffic,
and the results show that it has high detection accuracy.

\section{Method Overview}

We assume the data to be a sequence of \emph{interaction records}; each
record $\mathtt{r=(timestamp,id1,id2)}$ contains a timestamp and the IDs
of the two participants. For botnets of network computers,
a \emph{interaction record} corresponds to a network packet. 

We group \emph{interaction records} into windows based on their timestamps.
For all $k$'s, we denote by $\scrW_{k}$ the collection of \emph{interaction
records} in window $k$ and present the definition of the Social Interaction
Graph (SIG) for window $k$ as follows.

\begin{defi}
  (\emph{Social Interaction Graph}). Let $\scrE_{k}$ be an edge set such
  that $(i,j)\in \scrE_{k}$ if there exists at least one interaction
    record $\mathtt{r}\in\scrW_{k}$ whose participant IDs are $i$ and
  $j$. Then, the SIG $\scrG_{k}=\left(\scrV,\scrE_{k}\right)$ corresponding to
  $\scrW_{k}$ is an undirected graph whose vertex set $\scrV$ is the set of
  all nodes in the network and whose edge set is $\scrE_{k}$.
\end{defi}

On a notational remark, throughout the paper we will use $n$ to denote
the number of nodes in the network (cardinality of $\scrV$).

\begin{figure}[ht]
\begin{center}
\vspace{-0.3cm}
\includegraphics[width=8cm,height=6cm]{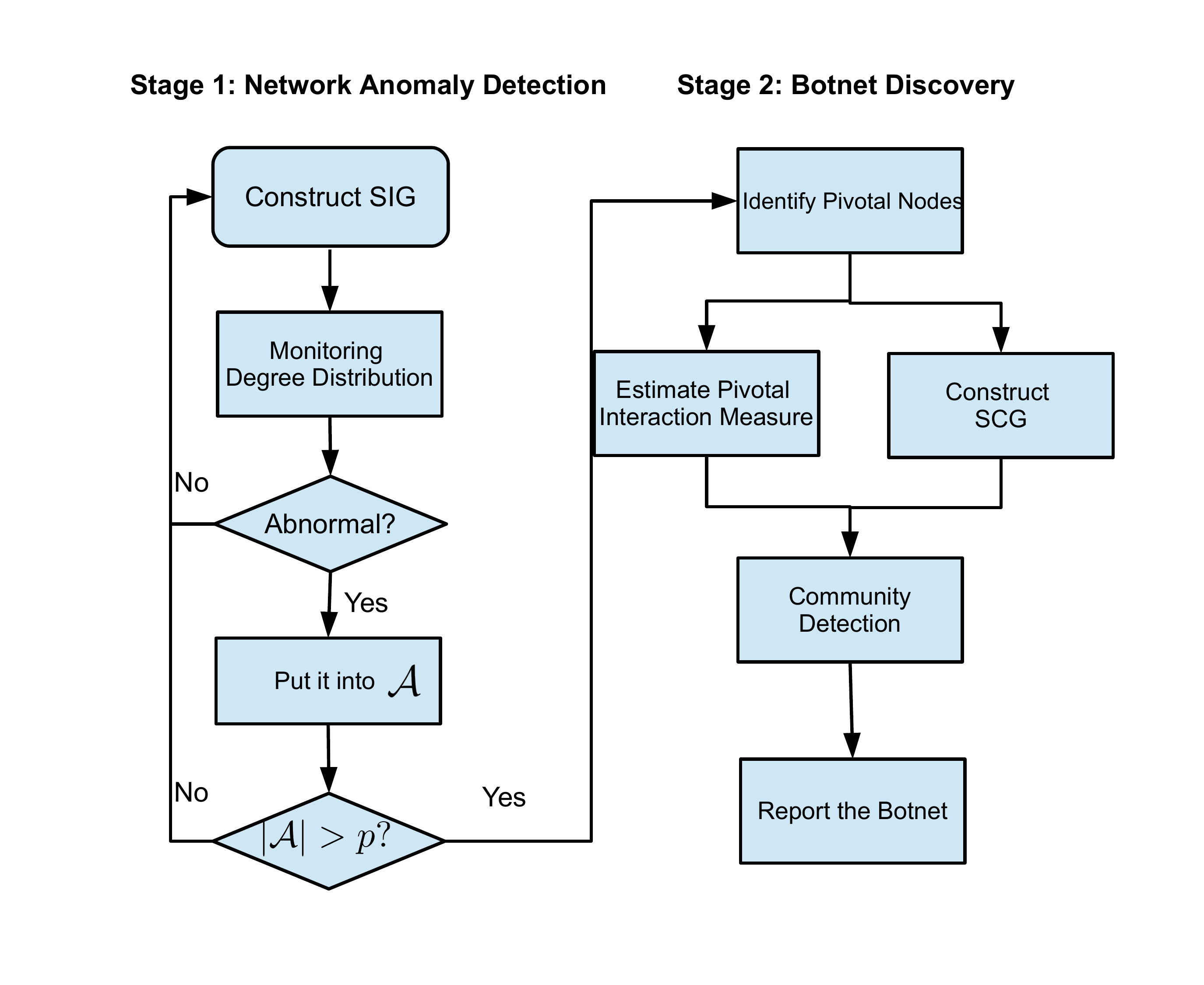}
\end{center}
\vspace{-0.5cm}
\caption{Overview of Our Method. \label{fig:method-overview}}
\end{figure}

Our method consists of a \emph{network anomaly detection} stage and a
\emph{botnet discovery} stage~(see Fig.~\ref{fig:method-overview}).  In
the \emph{network anomaly detection} stage, each SIG is evaluated with a
reference model and abnormal SIGs are stored into a pool $\calA$.  The
\emph{botnet discovery} stage is triggered whenever the size of the pool
$\calA$ is greater than a threshold $p$. A set of highly interactive
nodes, referred to as \emph{pivotal nodes}, are identified.  Both
botmaster and targets are very likely to be \emph{pivotal nodes} because
they need to interact with bots frequently. These interactions
correspond to C\&C traffic for botmasters and to attacking traffic for
targets.  In either case, the interactions between each bot and
\emph{pivotal nodes} should be correlated. To characterize this
correlation, we construct a \emph{Social Correlation Graph}~(SCG), whose
formal definition is in Section~\ref{sub:social-cor-graph}. We can
detect bots by detecting the community that has high interaction with
\emph{pivotal nodes} in the SCG. We propose a novel community detection
method based on a refined modularity measure. This modularity measure
uses information in SIGs, i.e., \emph{pivotal interaction measure} (see
Section~\ref{sub:refined-modularity}), to improve detection accuracy.

\section{Network Anomaly Detection}

As noted above, the goal of the \emph{network anomaly detection} stage
is to identify abnormal SIGs given some knowledge of what constitutes
``normal'' interactions between nodes.  A natural way is to monitor the
degree distributions of graphs and to compare them with appropriate
reference graph models. This paper focuses on the
Erd\"{o}s-R\'{e}nyi (ER) model, the most common type of random
graph models. Our approach, however, can be generalized to more types of
models. We apply \emph{composite hypothesis testing} to detect abnormal
graphs.

\subsection{Large Deviation Principle for ER Random Graphs}

First, we present a {\em Large Deviation Principle~(LDP)} for undirected
random graphs. Let $\scrG_{n}$ denote the space of all simple labeled
undirected graphs of $n$ vertices. For any graph $\scrG\in\scrG_{n}$,
let $\bd=\left(d_{1},\dots,d_{n}\right)$ denote the labeled degree
sequence of $\scrG$. Also let $m=\frac{1}{2}\sum_{j=1}^{n}d_{j}$ denote
the number of edges in $\scrG$. We assume that any two nodes are
connected by at most one edge, which means that the node degree in
$\scrG$ is less than $n$. For $0\leq i\leq n-1$, let
$h_{i}=\sum_{j=1}^{n}1\left(d_{j}=i\right)$ be the number of vertices in
$\scrG$ of degree $i$, where $1(\cdot)$ is the indicator
function. Henceforth, $\bh=\left(h_{0},\dots,h_{n-1}\right)$, a quantify
irrelevant to the ordering of vertices, will be referred to as the
\emph{degree frequency vector} of a graph $\scrG$. The empirical
distribution of the degree sequence $\bd$, defined by $\bmu^{(n)}$, is a
probability measure on $\mathbb N_{0}=\mathbb N\cup\left\{ 0\right\} $ that puts
mass $h_{i}/n$ at $i$, for $0\leq i\leq n-1$.

In the Erd\"{o}s-R\'{e}nyi model, $\scrG(n,p)$, the distribution of the
degree of any particular vertex $v$ is binomial. Namely,
$P(d_{v}=k)={n-1 \choose k} p^{k}\left(1-p\right)^{n-1-k},$ where $n$ is
the total number of vertices in the graph. It it well known that when
$n\ra\infty$ and $np$ is constant, the binomial distribution converges
to a Poisson distribution. Let $\beta=np$ denote the constant. Then in the
limiting case, the probability that the degree of a node equals $k$
is $p_{\beta,k}=\frac{\beta^{k}e^{-\beta}}{k!}$, which is
independent of the node label. Let
$\bp_{\beta}=\left(p_{\beta,0},\dots,p_{\beta,\infty}\right)$ be the
Poisson distribution viewed as a vector whose parameter is $\beta$.

Let $\mathbb P(\mathbb N_{0})$ be the space of all probability measures defined
on $\mathbb N_{0}$. We view any probability measure $\bmu\in\mathbb
P(\mathbb N_{0})$
as an infinite vector $\bmu=\left(\mu_{0},\dots,\mu_{\infty}\right)$.
Let $\calS=\left\{ \bmu\in\mathbb P\left(\mathbb N_{0}\right):\bar{\mu}:=\sum_{i=0}^{\infty}i\mu_{i}<\infty\right\} $
be the set of all probability measures on $\mathbb N_{0}$ with finite
mean. 
It is easy to verify that $\bp_{\beta}\in\calS$. Let $\bP_{n}$ denote
the Erd\"{o}s-R\'{e}nyi distribution on the space $\scrG_{n}$
with parameter $\beta/n$.

The so-called \emph{rate function}
$I:\calS\ra\left[-\infty,\infty\right]$ can be used to quantify the
deviations of $\bmu^{(n)}$ with respect to a random graph
model~(\cite{Mukherjee13,deze2}). For the ER model, \cite{Mukherjee13}
proposes the following rate function.

\begin{defi}
  \label{def:rate-er-model} For the ER model with parameter $\beta$ for
  its degree distribution, we could define the rate function
  $I_{ER}:\calS\ra\left[-\infty,\infty\right]$ as
\[
I_{ER}\left(\bmu;\beta\right)=D\left(\bmu\parallel\bp_{\beta}\right)+\frac{1}{2}\left(\bar{\mu}-\beta\right)+\frac{\bar{\mu}}{2}\log\beta-\frac{\bar{\mu}}{2}\log\bar{\mu},
\]
where $D\left(\bmu\parallel\bp_{\beta}\right)=\sum_{i}\mu_{i}\log\left(\frac{\mu_{i}}{p_{\beta,i}}\right)$
is the Kullback\textendash{}Leibler (KL) divergence of $\bmu$ with
respect to $\bp_{\beta}.$
\end{defi}
\cite{Mukherjee13} further establishes an LDP for $\bmu^{(n)}$ with this
rate function. In the interest of space, we will not provide a formal
statement of the LDP. Intuitively, when $n$ is large enough, the
empirical degree distribution behaves as
$\bP_{n}\left(\bmu^{(n)}\approx\bmu\right)\asymp
e^{-nI_{ER}(\bmu;\beta)}$.

\subsection{A Formal Anomaly Detection Test}

In this section, we consider the problem of evaluating whether a graph
$\scrG$ is normal, i.e., comes from the ER model with a certain set of
parameters~$\left(\calH_{0}\right)$.  Let $\bmu_{\scrG}$ be the
empirical degree distribution of the graph $\scrG$ and let
$I_{ER}\left(\bmu_{\scrG};\beta\right)$ (cf.
Def.~\ref{def:rate-er-model}) be the corresponding rate function. We
present the following statement of the generalized Hoeffding test for
this anomaly detection problem.
\begin{defi}
\label{def:mf-GHT}%
The Hoeffding test~\cite{hoef65} is to reject $\calH_{0}$
when $\scrG$ is in the set: 
\begin{equation}
S_{F}^{*}=\left\{ \scrG\mid I_{ER}\left(\bmu_{\scrG};\beta\right)\geq\lambda\right\} ,\label{eq:GHT}
\end{equation}
where $\lambda$ is a detection threshold. 

It can be shown that the Hoeffding test (\ref{eq:GHT}) satisfies the
Generalized Neyman-Pearson (GNP) criterion~\cite{deze2}.
\end{defi}

\section{Botnet Discovery}

The network anomaly detection technique in the previous section can only
report an alarm when a botnet exists. In order to learn more about the
botnet, we develop the \emph{botnet discovery} technique described in
this section. The first challenge for \emph{botnet discovery} is that a
single abnormal SIG is usually insufficient to infer complete
information about a botnet, including the botmasters and the bots in the
botnet. As a result, we monitor windows continuously and store all
abnormal SIGs in a pool $\calA$. The \emph{botnet discovery} stage is
triggered only when $|\calA|>p$.

\subsection{Identification of Pivotal Nodes}

We assume a sequence of abnormal SIGs $\calA=\left\{
  \scrG_{1},\dots,\scrG_{|{\cal A}|}\right\} $. 
Detecting bots directly is non-trivial. Instead, detecting the leaders
(botmasters) or targets is much simpler because they are more interactive
than normal nodes. Botmasters need to ``command and control'' their
bots in order to maintain the botnet, and bots actively interact with
victims in typical DDoS attacks. Both leaders and targets, henceforth
referred to as \emph{pivotal nodes}, are highly interactive. Let $G_{k}^{ij}$
be an indicator of edge existence between node $i$ and $j$ in $\scrG_{k}$.
Then, for $i=1,\dots,n$, 
\begin{equation}
e_{i}=\frac{1}{\left|\calA\right|}\sum_{k=1}^{|\calA|}\sum_{j=1}^{n}G_{k}^{ij}\label{eq:interaction-in-A}
\end{equation}
represents the amount of interaction of node $i$ with all other nodes
in $\calA$. Henceforth, $e_{i}$ is referred to as the \emph{total
interaction measure} of node $i$. We present the following definition
of \emph{pivotal nodes}. 
\begin{defi}
  (\emph{Pivotal nodes}). We define the set of \emph{pivotal nodes}
  $\calN=\left\{ i:e_{i}>\tau\right\} $, where $\tau$ is a threshold.
\end{defi}
After identifying \emph{pivotal nodes}, the problem is equivalent to
detecting the community associated with \emph{pivotal nodes}.

\subsection{Botnet Discovery}

\subsubsection{Construction of the Social Correlation Graph\label{sub:social-cor-graph}}

Compared to similar approaches in community detection, e.g., the
leader-follower algorithm~\cite{Shah2010}, our method takes advantage of
not only temporal features (SIG) but also correlation
relationships. These relationships are characterized using a graph,
whose definition is presented next.

For $i=1,\dots,n$, let variable $X_{i}$ represent the number of
\emph{pivotal nodes} in $\calN$ that node $i$ has interacted with. Let
$\rho(X_{i},X_{j})$ be the sample Pearson correlation coefficient
between two random variables $X_{i}$ and $X_{j}$. In addition, if the
sample standard deviation of either $X_{i}$ or $X_{j}$ equals zero, we
let $\rho\left(X_{i},X_{j}\right)=0$ to avoid division by zero.
\begin{defi}
  \emph{\label{def:cor-graph}(Social Correlation Graph}). The Social
  Correlation Graph (SCG) $\scrC=\left(\scrV,\scrE_{c}\right)$ is an
  undirected graph with vertex set $\scrV$ and edge set
  $\scrE_{c}=\left\{
    \left(i,j\right):\left|\rho\left(X_{i},X_{j}\right)\right|>\tau_{\rho}
  \right\} $, where $\tau_{\rho}$ is a threshold.
\end{defi}
Because the behaviors of the bots are correlated, they are more likely
to be connected to each other in the SCG. Our problem is to find an
appropriate division of the SCG to separate bots and normal nodes.  Our
criterion for ``appropriate'' is related to the well-known concept of
\emph{modularity} in community
detection\emph{~\cite{Newman2004Fast,Newman2004,Newman2006}}.

\subsubsection{Modularity-based Community Detection}

The problem of community detection in a graph amounts to dividing the
vertices of a given graph into non-overlapping groups such that
connections within groups are relatively dense while those between
groups are sparse~\cite{Newman2004}.

The \emph{modularity} for a given subgraph is defined to be the fraction
of edges within the subgraph minus the expected fraction of such edges
in a randomized null model. Although it was proposed as the stopping
criterion of a method, this measure later inspired a broad range of
community detection methods named \emph{modularity-maximization}
methods.

We consider the simple case when there is only one botnet in the
network.  As a result, we want to divide the nodes into two groups, one
for bots and one for normal nodes. Suppose that $s_{i}$ is variable such
that $s_{i}=1$ if node $i$ is a bot and $s_{i}=-1$ otherwise. Let
$d_{i}^{c}$ be the degree of node $i$ in SCG
$\scrC=\left(\scrV,\scrE_{c}\right)$ for $i=1,\dots,n$ and let
$m^{c}=\frac{1}{2}\sum_{i}d_{i}^{c}$ be the edge number of $\scrC$. For
a partition specified by $\bs=\left(s_{1},\dots,s_{n}\right)$, its\emph{
  modularity} is defined as in \cite{Newman2004}
\begin{equation}
Q\left(\bs\right)=\frac{1}{2m^{c}}\sum_{i,j=1}^{n}\left(A_{ij}-N_{ij}\right)\delta\left(s_{i},s_{j}\right),\label{eq:trad_modularity}
\end{equation}
where
$\delta\left(s_{i},s_{j}\right)=\frac{1}{2}\left(s_{i}s_{j}+1\right)$ is
an indicator of whether node $i$ and node $j$ are of the same type.
$A_{ij}=1\left(\left|\rho\left(X_{i},X_{j}\right)\right|>\tau_\rho\right)$
is an indicator of the adjacency of node $i$ with node $j$. $N_{ij}$ is
the \emph{expected number of edges} between $ $node $i$ and node $j$ in
a null model. The selection of the null model is empirical, but the most
common choice by far is the \emph{configuration model}~\cite{Molloy1995}
in which $N_{ij}=\frac{d_{i}^{c}d_{j}^{c}}{2m^{c}}$. The optimal
division of vertices should maximize the modularity measure
(\ref{eq:trad_modularity}).

\subsubsection{Refined Modularity\label{sub:refined-modularity}}

We introduce two refinements to the \emph{modularity} measure to make it
suitable for botnet detection. First, intuitively, bots should have
strong interactions with \emph{pivotal nodes} and normal nodes should have
weak interactions. We want to maximize the difference. As a result, our
objective considers nodes' interaction to the pivotal nodes. Let
\begin{equation}
r_{i}=\frac{1}{\left|\calA\right|}\sum_{k=1}^{|\calA|}\sum_{j\in\calN}e_{i}G_{k}^{ij}\label{eq:interaction-to-pivot-nodes}
\end{equation}
denote the amount of interaction between node $i$ and \emph{pivotal nodes}.
We refer to $r_{i}$ as \emph{pivotal interaction measure} of
node $i$. Then, $\sum_{i}r_{i}s_{i}$ quantifies the difference between
the \emph{pivotal interaction measure} of bots and that of normal
nodes. A natural extension for the modularity measure is to include an
additional term to maximize $\sum_{i}r_{i}s_{i}$.

Second, the \emph{modularity} measure is criticized to suffer from low
resolution, namely it favors large communities and ignores small
ones~\cite{Santo2007,Lancichinetti2011}.  The botnet, however, could
possibly be small. To address this issue, we introduce a regularization
term for the size of botnets. It is easy to obtain that
$\sum_{i}\mathbf{1}\left(s_{i}=1\right)=\sum_{i}\frac{s_{i}+1}{2}$ is
the number of detected bots. Thus, our refined \emph{modularity }measure
is
\begin{eqnarray}
Q_{d}\left(\bs\right) & = & \frac{1}{2m^{c}}\sum_{i,j\in \scrV}\left(A_{ij}-\frac{d_{i}^{c}d_{j}^{c}}{2m^{c}}\right)s_{i}s_{j}\nonumber \\
 &  & +w_{1}\sum_{i}r_{i}s_{i}-w_{2}\sum_{i}\frac{s_{i}+1}{2}\label{eq:synthetic-modularity}
\end{eqnarray}
where $w_{1}$ and $w_{2}$ are appropriate weights.

The two modifications also influence the results of isolated nodes with
degree $0$, which possibly exist in SCGs. By Def.~\ref{def:cor-graph}, a
node is isolated if its sample deviation is zero or its correlations
with other nodes are small enough. The placement of isolated nodes,
however, does not influence the traditional modularity measure,
resulting in arbitrary community detection
results~\cite{Newman2004}. This limitation is addressed by the two
additional terms. If node $i$ is isolated and $r_{i}=0$, then $s_{i}=-1$
in the solution because of the regularization term
$w_{2}\sum_{i}\frac{s_{i}+1}{2}$. On the contrary, if $r_{i}$ is large
enough, $s_{i}=1$ in the solution because of the term
$w_{1}\sum_{i}r_{i}s_{i}$.

\subsection{Relaxation of the Optimization Problem}

The modularity-maximization problem has been shown as being
NP-complete~\cite{Agarwal2008a,Brandes2008}. The existing algorithms for
this problem can be broadly categorized into two types: $(i)$ heuristic
methods that solve this problem directly~\cite{Duch2005}, and $(ii)$
mathematical programming methods that relax it into an easier problem
first~\cite{Agarwal2008a,chan2011convex}. We follow the second route
because it is more rigorous.

We define the modularity matrix $\bM=\left\{ M_{ij}\right\}
_{i,j=1}^{n}$, where
$M_{ij}=\frac{A_{ij}}{2m^{c}}-\frac{d_{i}^{c}d_{j}^{c}}{\left(2m^{c}\right)^{2}}$.
Let $\bs=\left(s_{1},\dots,s_{n}\right)$ and
$\br=\left(r_{1},\dots,r_{n}\right)$, then the modularity-maximization
problem becomes
\begin{align}
\max\qquad & \bs^{'}\bM\bs+\left(w_{1}\br^{'}-\frac{w_{2}}{2}\mathbf{1}^{'}\right)\bs\label{eq:org_opt}\\
s.t.\qquad & s_{i}^{2}=1,\qquad \forall i. \nonumber 
\end{align}
To make the objective function concave, we introduce a negative multiple
of $\bs^{'}\bI\bs$ \cite{chan2011convex}, leading to:
\begin{align}
\max\qquad & \bs^{'}\left(\bM-\sigma \bI\right)\bs+\left(w_{1}\br^{'}-\frac{w_{2}}{2}\mathbf{1}^{'}\right)\bs\label{eq:revised_opt}\\
s.t.\qquad & s_{i}^{2}=1,\qquad \forall i, \nonumber 
\end{align}
where $\sigma$ is a positive scalar. Notice that the objective of
(\ref{eq:revised_opt}) is equivalent to that of (\ref{eq:org_opt})
because $\bs^{'}\bI\bs=n s_{i}^{2}=n$ is ensured by the
constraint. We can choose $\sigma$ large enough so that
$\bM-\sigma \bI$ is negative definite. This modification induces no
extra computational cost. Although the feasible domain of the revised
problem is still non-convex, the objective is concave
now. (\ref{eq:revised_opt}) is a typical non-convex Quadratically
Constrained Quadratic Programming (QCQP)~\cite{Aspremont2003}. Let
$\bS=\bs\bs^{'}$, $\bP_{0}=\bM-\sigma\bI$, and
$\bq_{0}=w_{1}\br-\frac{w_{2}}{2}\mathbf{1}$. We can relax problem
(\ref{eq:revised_opt}) to
\begin{equation}
\begin{array}{rl}
\max & \Tr\left(\bS\bP_{0}\right)+\bq_{0}^{'}\bs \\
s.t. & \begin{bmatrix}
\bS & \bs\\
\bs^{'} & 1
\end{bmatrix} \succeq 0,\\
& S_{ii} =1,\qquad \forall i.
\end{array}
\label{eq:SDP_relax}
\end{equation}
The problem above is a {\em Semidefinite Programming problem (SDP)} and
produces an upper bound on the optimal value of the original problem~\cite{Aspremont2003}. It
is well known that SDP is polynomially solvable and many solvers
(CSDP~\cite{Borchers1999}, SDPA~\cite{fujisawa1995sdpa}) are
available. 

\subsubsection{Randomization}

The SDP relaxation (\ref{eq:SDP_relax}) provides an optimal solution
together with an upper bound on the optimal value of problem
(\ref{eq:revised_opt}). However, the solution of the SDP relaxation
(\ref{eq:SDP_relax}) may not be feasible for the original problem
(\ref{eq:revised_opt}). To generate feasible solutions we use a
randomization technique.

If $\left(\bS^{*},\bs^{*}\right)$ is the optimal solution of the relaxed
problem, then $\bS^{*}-\bs^{*}\bs^{*'}$ can be interpreted as a
covariance matrix. If we pick $\bx=\left(x_{1},\dots,x_{n}\right)$ as
a Gaussian random vector with
$\bx\sim\scrN(\bs^*,\bS^{*}-\bs^{*}\bs^{*'})$, then $\bx$ ``solves''
the non-convex QCQP in (\ref{eq:revised_opt}) ``on average'' over this
distribution. 
As a result, we can draw samples $\bx$ from this normal distribution and
simply obtain feasible solutions by taking $\hat{\bx}=\sgn(\bx)$. We
sample 10,000 points and pick the point that maximizes
$f(\bx)=\bx^{'}\left(\bM-\sigma\bI\right)\bx+\left(w_{1}\br-\frac{w_{2}}{2}\mathbf{1}\right)^{'}\bx$.

\section{Experimental Results}

In this section, we apply our \emph{network anomaly detection} approach
to real-world traffic. Meanwhile, we also compare the performance of our
\emph{botnet discovery} approach, a modularity-based community detection
technique, with existing community detection techniques.

\subsection{Description of Dataset}

In this paper, we mix some real-world botnet traffic with some real-world
background traffic. For the real-world botnet traffic, we use the
``DDoS Attack 2007'' dataset by the Cooperative Association for
Internet Data Analysis (CAIDA)~\cite{url-caida-ddos}. It includes
traces from a Distributed Denial-of-Service (DDoS) attack on August
4, 2007. The DDoS attack attempts to block access to the targeted
server by consuming computing resources on the server and by consuming
all of the bandwidth of the network connecting the server to the Internet.

The total size of the dataset is $21$ GB and the dataset covers about one
hour (20:50:08 UTC to 21:56:16 UTC). These dataset only contains attacking
traffic to the victim; all other traffic, including the C\&C traffic, has
been removed by the creator of the dataset. The dataset consists of
two parts. The first part is the traffic when the botnet initiates
the attack (between 20:50~UTC and 21:13~UTC). In the initiating
stage, the bots probe whether they can reach the victim in order to
determine the set of nodes that should participate in the attack.
The traffic of the botnet during this period is small, thus, it is
very challenging to detect it using only network load. The second
part is the attack traffic which starts around 21:13 UTC when the
network load increases rapidly (within a few minutes) from about $200$
Kb/s to about $80$ Mb/s. With this significant change of transmission
rate, it is trivial to detect botnets when the attack starts (after
21:13~UTC). In this paper, we select a $5$-minutes segment from
the first part, i.e., during the time when the botnet initiates the
attack. The total number of bot IP addresses in the selected traffic
is $136$.

For the background traffic, we use trace $6$ in the University of Twente
traffic traces data repository ($\mathtt{simpleweb}$)~\cite{eemcs17829}.
This trace was measured in a $100$ Mb/s Ethernet link connecting
an educational organization to the Internet. This is a relatively
small organization with around $35$ employees and a little over $100$
students working and studying at this site (the headquarters
of this organization). There are $100$ workstations at this location
which all have $100$ Mbit/s LAN connection. The core network consists
of a $1$ Gbit/s connection. The recordings took place between the
external optical fiber modem and the first firewall. The measured
link was only mildly loaded during this period. The background traffic
we choose lasts for $3,600$ seconds. The botnet traffic is mixed with
background traffic between $2,000$ and $2,300$ seconds.

\subsection{Results of Network Anomaly Detection\label{sub:res-network-ano-det}}

\begin{figure}[h]
\begin{centering}
\includegraphics[width=0.9\linewidth]{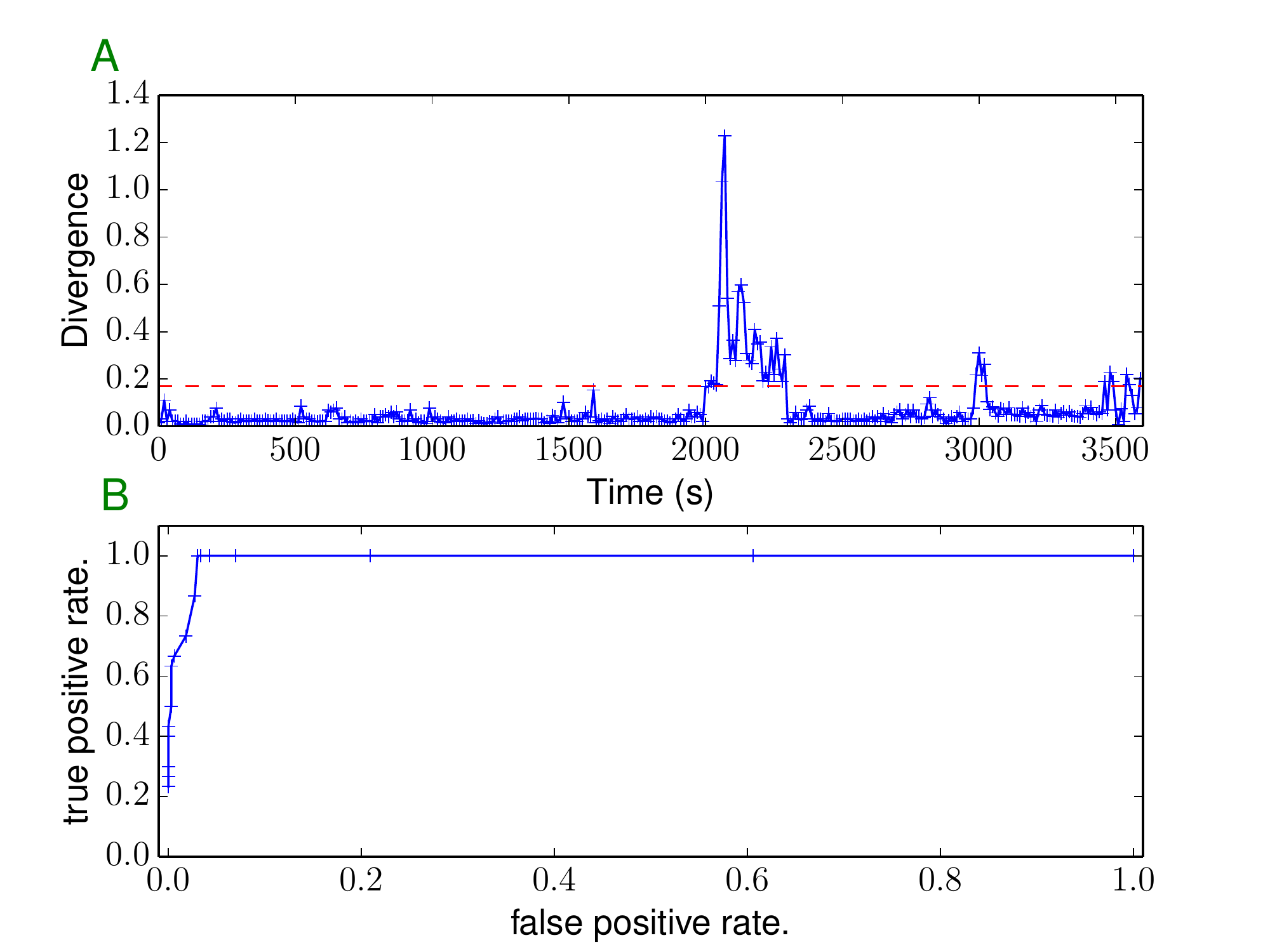}
\par\end{centering}
\caption{Figure~A shows the rate function value
  $I_{ER}(\bmu_{i};\hat{\beta})$ for each window $i$. The
  $x$-axis plots the starting time of each window.  The background traffic
  lasts for 3,600 seconds and the botnet traffic is added between 2,000
  and 2,300 seconds. Figure~B shows the ROC curve.  The $x$-axis plots
  the false alarm rate and the $y$-axis the true positive
  rate.\label{fig:caida-simpleweb-det-roc-combine}}
\vspace{-0.3cm}
\end{figure}

We divide the mixed traffic into $10$-second windows and create a
sequence of $360$ SIGs.
Fig.~\ref{fig:caida-simpleweb-det-roc-combine}-A shows the detection
results. The blue ``$\text{+}$'' markers indicate the value of
$I_{ER}(\bmu_{i};\hat{\beta})$ for each window $i$,
$i=1,\dots,360$, where $\bmu_{i}$ is the empirical degree distribution
of SIG $i$ and $\hat{\beta}$ is estimated from the SIGs created using
only background traffic. The red dash line shows the threshold
$\lambda=0.18$, which can be set to constrain the false alarm rate below
a desirable value. According to rule (\ref{eq:GHT}), there are 36
abnormal SIGs, namely $\left|\calA\right|=36$. There are 30 SIGs that
have botnet traffic and 29 SIGs are correctly identified. SIG no. 20
corresponding to the time range $[2000s,2010s]$ is missed. Being the
start of the botnet traffic, this range has very low botnet activity,
which may explain the miss-detection. In addition, there are two groups
of false alarms---3 false alarms around 3,000s and 4 false alarms around
3,500s. Fig.~\ref{fig:caida-simpleweb-det-roc-combine}-B shows the
Receiver Operating Characteristic (ROC) curve of the detection rule
(\ref{eq:GHT}).

\subsection{Results of Botnet Discovery}

\begin{figure}
\begin{centering}
\includegraphics[width=0.8\columnwidth]{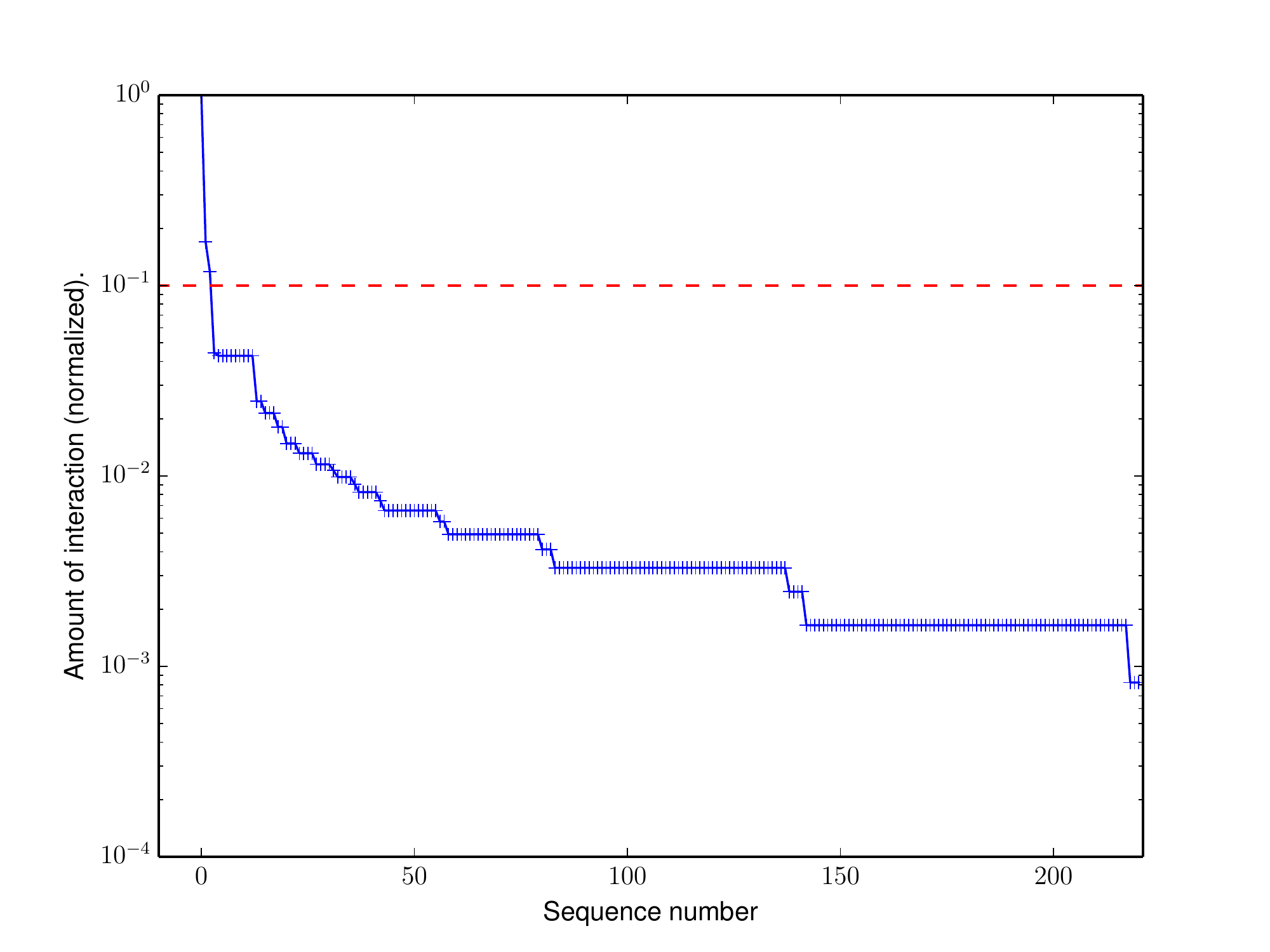}
\par\end{centering}

\caption{Sorted amount of interaction in $\calA$ defined by (\ref{eq:interaction-in-A}).
$y$-axis is in log-scale. \label{fig:total-interaction-in-A}}
\end{figure}

The \emph{botnet discovery} stage aims to identify bots based on the
information in $\calA$. The first step is to identify a set of pivotal
nodes. Recall that the \emph{total interaction measure} $e_{i}$ in
(\ref{eq:interaction-in-A}) quantifies the amount of interaction in
$\calA$ of node $i$ with other nodes. The set of pivotal nodes is
$\calN=\left\{ i:e_{i}>\tau\right\} $, where $\tau$ is a prescribed
threshold. Let $e_{max}$ be the maximum \emph{total interaction measure}
of all nodes and $\scrS_{e}^{Norm}=\left\{
  e_{i}/e_{max}:i=1,\dots,n\right\} $ be the normalized set of
\emph{total interaction measures}. Fig.~\ref{fig:total-interaction-in-A}
plots $\scrS_{e}^{Norm}$ in descending order and in log-scale for the
$y$-axis.  Each blue ``$\text{+}$'' marker represents one node. The blue
curve in Fig.~\ref{fig:total-interaction-in-A}, being quite steep,
clearly indicates the existence of influential pivotal nodes. The red dash
line in Fig.~\ref{fig:total-interaction-in-A} plots the selected
threshold $\tau$, which results in $3$ pivotal nodes. Only one pivotal node
belongs to the botnet. The other two pivotal nodes are active normal
nodes. These two falsely detected pivotal nodes correspond to the two
false-alarm groups described in Section~\ref{sub:res-network-ano-det}.

\begin{figure}
\begin{centering}
\includegraphics[width=7cm]{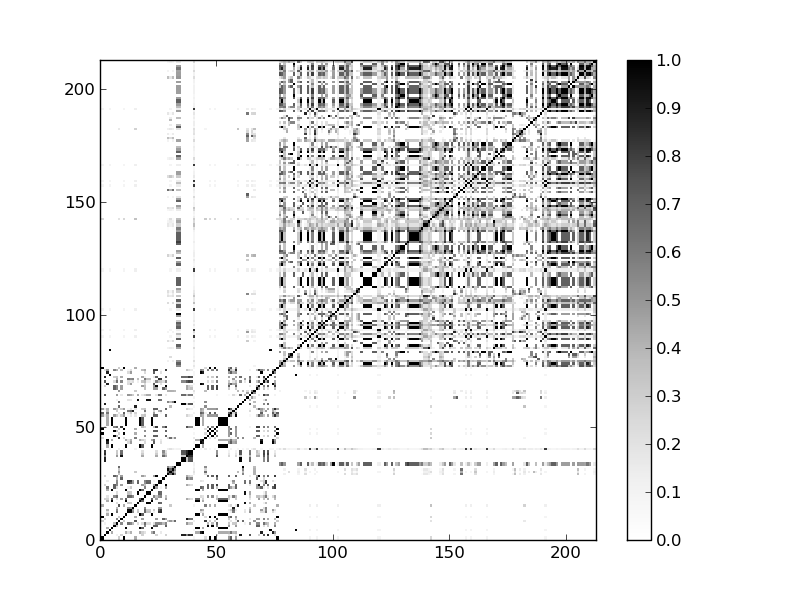}
\par\end{centering}
\caption{The correlation matrix $\left[\rho\left(X_{i},X_{j}\right)\right]_{i\in \scrV_{p},j\in \scrV_{p}}$
.
\label{fig:cor-mat}}
\end{figure}

Our dataset has $396$ nodes, including $136$ bots and $260$ normal
nodes. Among the $396$ nodes, only $213$ nodes have positive sample
standard deviations. Let $\scrV_{p}$ be the set of all nodes with
positive sample standard deviations, Fig.~\ref{fig:cor-mat} plots the
correlation matrix of these nodes. We can easily observe two groups from
Fig.~\ref{fig:cor-mat}.

\begin{figure*}[t]
\begin{centering}
\includegraphics[width=0.9\textwidth,height=7.5cm]{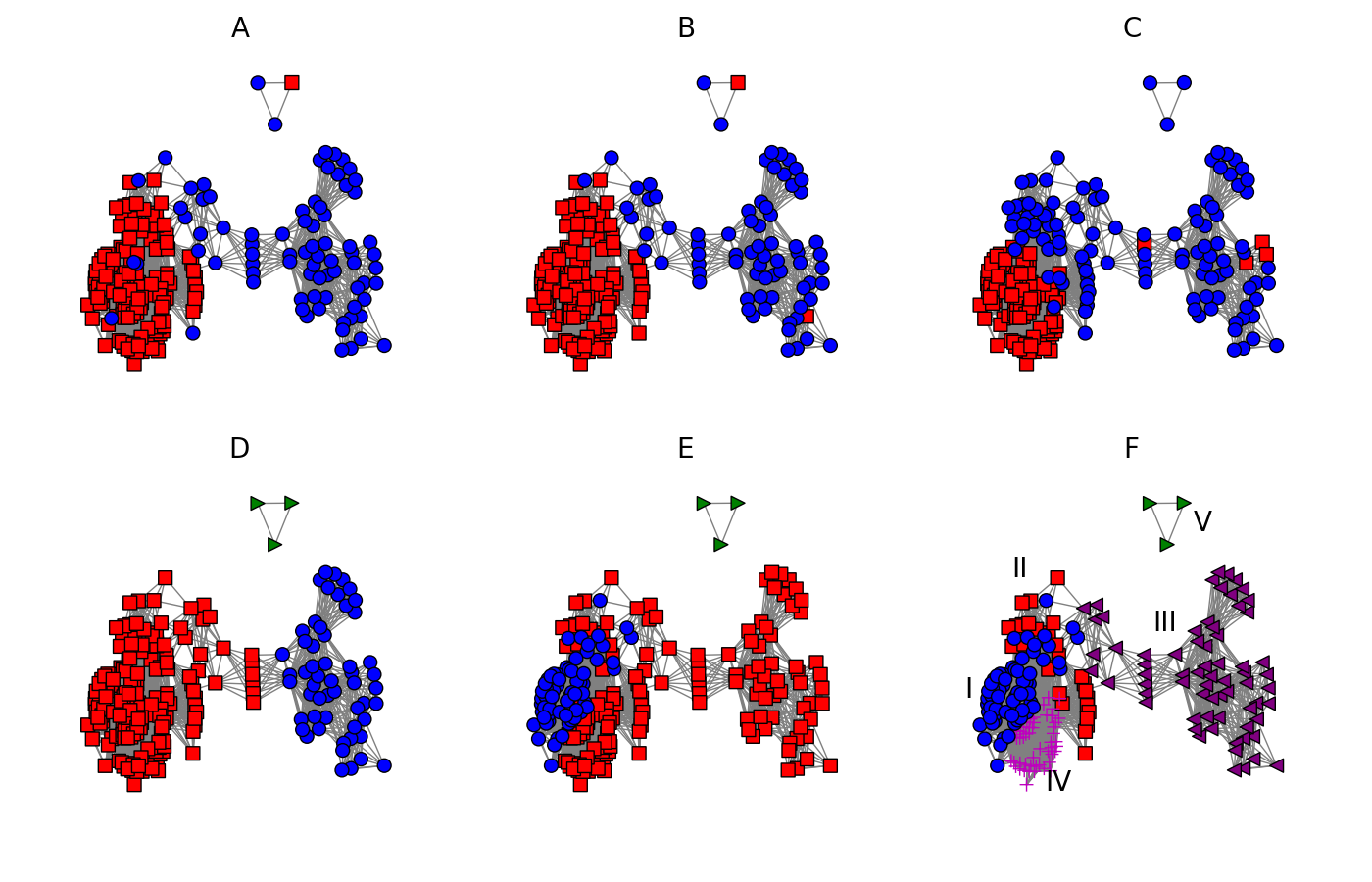}
\par\end{centering}
\vspace{-1cm}
\caption{Comparison of different community detection techniques on
  SCG. Fig.  A shows the ground-truth communities of bots and normal
  nodes. Fig.  B is the result of our botnet discovery approach. Fig. C
  is the result of the vector programming method proposed by Agarwal
  et~al.~\cite{Agarwal2008a}.  Fig. D is the
  result of the walktrap method~\cite{Pascal2005} with three
  communities. Fig. E is the result of Newman's leading eigenvector
  method~\cite{Newman2006} with 3 communities.  Fig. F is the result of
  the leading eigenvector method with 5
  communities. In figure A-C, red squares are bots and blue circles are
  normal nodes. In figure D-F, red squares indicate the group with
  highest average \emph{pivotal interaction measure}, while blue circles
  indicate the group with the lowest
  one. \label{fig:com-det-comparison}}
\end{figure*}
We calculate the SCG $\scrC$ using Def.~\ref{def:cor-graph} and
threshold $\tau_{\rho}=0.3$. In the SCG $\scrC$, there are 191 isolated
nodes with degree zero. The subgraph formed by the remaining 205 nodes
has two connected
components~(Fig.~\ref{fig:com-det-comparison}-A). Fig.~\ref{fig:com-det-comparison}-A
plots normal nodes as blue circles and bots as red squares. Although the
bots and the normal nodes clearly belong to different communities, the
two communities are not separated in the narrowest part of the
graph. Instead, the separating line is closer to the bots.

We apply our \emph{botnet discovery} method to $\scrC$. The result
(Fig.~\ref{fig:com-det-comparison}-B) is very close to the ground
truth~(Fig.~\ref{fig:com-det-comparison}-A).  As comparison, we also
apply other community-detection methods to the 205-node subgraph.

The first method is the vector programming method proposed by Agarwal et
al.~\cite{Agarwal2008a}, which is a special case of our method in which
$w_{1}=0$ and $w_{2}=0$. This approach, however, misses a number of
bots~(\ref{fig:com-det-comparison}-C).

The second method is the $\mathtt{walktrap}$ method by Pascal et
al.~\cite{Pascal2005,csardi2006igraph}, which defines a distance measure
for vertices based on a random walk and applies hierarchical
clustering~\cite{ward1963hierarchical}.  When the desirable number of
communities, a required parameter, equals to two, the method outputs the
two connected components, a reasonable yet useless result for botnet
discovery. To make the results more meaningful, we use
$\mathtt{walktrap}$ to find three communities and ignore the smallest
one that corresponds to the smaller connected components~(right
triangles in Fig.~\ref{fig:com-det-comparison}-D).  The community with a
higher mean of pivotal interaction measure is detected as botnet, and
the rest of the nodes are labeled as normal. The $\mathtt{walktrap}$
method separates bots and normal nodes in the narrowest part of the
graph, a reasonable result from the perspective of community
detection~(Fig.~\ref{fig:com-det-comparison}-D).  However, a comparison
with the ground-truth reveals that a lot of normal nodes are falsely
reported as bots.

The third method is the Newman's leading eigenvector
method~\cite{Newman2006,csardi2006igraph}, a classical modularity-based
community detection method. This method calculates the eigenvector
corresponding to the second-largest eigenvalue of the modularity matrix
$\bM$, namely the \emph{leading eigenvector}, and lets solution $\bs$ be
the sign of the \emph{leading eigenvector}.  The method can be
generalized for detecting multi-communities~\cite{Newman2006}. Similar
to the $\mathtt{walktrap}$ method, the leading eigenvector method
reports two connected components as results when the desirable community
number is two. We also use this method to find three communities and
ignore the smallest one. Again, the community with higher mean of
\emph{pivotal interaction measure} is detected as the botnet.

Different from previous methods, the eigenvector method makes completely
wrong prediction of the botnet. The community whose majority are bots
(blue circles in Fig.~\ref{fig:com-det-comparison}-E) is wrongly
detected as the normal part and the community formed by the rest of the
nodes is wrongly detected as the botnet. Despite being part of the real
botnet, the community of blue circles in
Fig.~\ref{fig:com-det-comparison}-E actually has lower mean of
\emph{pivotal interaction measure}, i.e., less overall communication
with pivotal nodes.

After dividing the SCG $\scrC$ into five communities using the leading
eigenvector approach for multi-communities~\cite{Newman2006}, we observe
that the botnet itself is heterogeneous and divided into three groups.
Both the group with the highest mean of \emph{pivotal interaction
  measure} (Group II in Fig.~\ref{fig:com-det-comparison}-F) and the
group with the lowest mean (Group I in
Fig.~\ref{fig:com-det-comparison}-F) are part of the botnet.

Because of the heterogeneity, some groups of the botnet may be
misclassified.  On the one hand, the leading eigenvector method wrongly
separates Group I from the rest as a single community, and merges Group
II \& IV with the normal part (Group III). Because Group I has the lowest
\emph{pivotal interaction measure}, it is wrongly detected as normal,
causing Group II, III, IV to be detected as the botnet. On the other
hand, the vector programming method wrongly detects a lot of nodes in
Group II, which should be bots, as normal nodes.

By taking the \emph{pivotal interaction measure} into consideration, the
misclassification can be avoided. In our formulation of refined
modularity (\ref{eq:synthetic-modularity}), the term
$w_{1}\sum_{i}r_{i}s_{i}$ maximizes the difference of the \emph{pivotal
  interaction measure} of the botnet and that of the normal part. Owing
to this term, our method makes little mistake for nodes in Group II
since they have high \emph{pivotal interaction measure}s.





\section{Conclusion}

In this paper, we propose a novel method of botnet detection that
analyzes the social relationships, modeled as Social Interaction Graphs
(SIGs) and Social Correlation Graphs (SCGs), of nodes in the network.
Compared to previous methods, our method has following novelties. First,
our method applies social network analysis to botnet
detection and can detect botnets with sophisticated C\&C channels. Second, our
method can be generalized to more types of networks, such as email networks
and biological networks~\cite{newman2002email,newman2009networks}.  Third,
we propose a refined modularity measure that is suitable for botnet
detection. The refined modularity also addresses some limitations of
\emph{modularity}. 

\bibliographystyle{IEEEtran}



\bibliography{allerton-social.bbl}

\end{document}